\documentclass[preprints,article,accept,moreauthors,pdftex]{Definitions/mdpi} 

\firstpage{1} 
\pubvolume{1}
\issuenum{1}
\articlenumber{0}
\pubyear{2021}
\copyrightyear{2021}
\datereceived{} 
\dateaccepted{} 
\datepublished{} 
\hreflink{https://doi.org/} 

\Title{Polarization Tomography with Stokes Parameters}

\TitleCitation{Polarization Tomography with Stokes Parameters}

\Author{Lawrence Rudnick $^{1,}$*$^{,\dagger}$ \orcidA{}, Debora Katz $^{2}$ and Lerato Sebokolodi $^{3}$ \orcidB{}}

\AuthorNames{Lawrence Rudnick, Debora Katz and Lerato Sebokolodi}

\AuthorCitation{Rudnick, L.; Katz, D.; Sebokolodi, L.}

\address{%
$^{1}$ \quad {University} 
of Minnesota, Minnesota Institute for Astrophysics, School of Physics and Astronomy, {Minneapolis, MN 55455, USA} 
\\
$^{2}$ \quad U.S. Patent and Trademark {Office}, Alexandria, VA 22314, USA 
; deborakatz@yahoo.com\\
$^{3}$ \quad  Rhodes {University}, Department of Physics and Electronics, Faculty of Science 
, {Drosty Rd, Grahamstown 6139, South Africa} 
; mll.sebokolodi@gmail.com
}

\corres{Correspondence: larry@umn.edu; Tel.: +1-612-624-3396}

\firstnote{Current address: Minnesota Institute for Astrophysics, University of Minnesota, 116 Church St., SE, Minneapolis, MN 55455, USA} 

\abstract{We present a simple but powerful technique for the analysis of polarized emission from radio galaxies and other objects. It is based on the fact that images of Stokes parameters often contain considerably more information than is available in polarized intensity and angle maps. In general, however, the orientation of the Stokes parameters will not be matched to the position angles of structures in the source.  Polarization tomography, the technique presented in this paper, consists of making a series of single linear Stokes parameter images, $S(\rho)$, where each image is rotated by an angle $\rho$ from the initial orientation of Q and U.  Examination of these images, in a series of still frames or a movie, reveals often hidden patterns of polarization angles, as well as structures that were obscured by the presence of overlapping polarized emission. We provide both cartoon examples and a quick look at the complex polarized structure in Cygnus A.}


\begin{document}

\section{Introduction}
Images of polarized intensity often contain more details than those of total intensity, because the fractional polarization can vary across a structure.  The fractional polarization is dependent on both the ordering of magnetic fields on microscopic scales, as well as larger-scale destructive and constructive combinations of polarized emission with different polarization angles.  Even where the  polarized intensity $P$ is constant across a structure, the polarization angle $\chi$ may vary, so these two images must be examined simultaneously.   Alternatively, all of this same information is present in the Stokes $Q$ and $U$ images, from which $P$ and $\chi$ are derived, according to:
\begin{equation}
    P = \sqrt(Q^2+U^2),
\end{equation}
and
\begin{equation}
    \chi = 0.5 \times \arctan(U,Q)
\end{equation}

Visualizing the polarization angle structures is difficult, and subtle gradients and other structures, {reflecting variations in the intrinsic magnetic field directions,} are often difficult to see.   The Stokes $Q$ and $U$ images can often reveal these behaviors, but their fixed orientation with respect to the sky may not be optimum.  Polarization tomography, introduced by \citet{ptomo}, and presented here, circumvents this limitation.

{More precisely, for experienced radio polarimetrists, polarization tomography works on Q($\lambda^2_a$), U(($\lambda^2_a$), where $\lambda^2_a$ is the wavelength at which the images are made.   If Faraday rotation is present (and it always is),  the structures which are revealed in the tomography maps will reflect both variations in the intrinsic magnetic field directions {\textbf and} in the Faraday medium.
If, on the other hand, Faraday rotation has been removed, then $\lambda^2_a = 0$, and the tomography structures will reflect only variations in the intrinsic magnetic field directions. }

Increasingly, the very broad bandwidths now available at the VLA, ASKAP, and MeerKAT are revealing the presence of more than one Faraday component along the line of sight, using either Faraday synthesis (\citet{rmsynth}) or Q,U fitting (\citet{farns}, \citet{osull}). Q and U are then actually functions of both $\lambda^2$ and Faraday depth $\phi$.  In this case, polarization tomography is best applied to maps corresponding separately to each single Faraday depth $\phi_i$, and de-rotated to $\lambda^2=0$.  We assume that simplification for the rest of this paper.

Although this technique was developed decades ago, it has not been in use.  Now that exquisitely complex polarization maps are becoming available (e.g., \citet{cyga}), it is timely for polarization tomography to become a standard analysis tool.

\section{Method}
We create a series of images $S(\rho)$ where
\begin{equation}
    S(\rho)= \cos{2\rho}\times Q + \sin{2\rho} \times U,
\end{equation}
where $\rho$ varies from 0 to 90$^{\circ}$, or, to improve visualization, from 0 to 180$^{\circ}$.  Since $Q = P\cos(2\chi)$ and $U = P\sin(2\chi)$, this is equivalent to
\begin{equation}
    S(\rho)=\cos{2\rho}\times P\cos(2\chi) + \sin{2\rho} \times P\sin(2\chi) = P cos(2(\rho-\chi)).
\end{equation}

Thus, a structure with polarization angle $\chi_0$ will appear most strongly in the S($\rho$) image where $\rho=\chi_0$ and will appear most strongly negative where $\rho-\chi_0 = 90^{\circ}$. Most importantly \emph{{it will be completely absent}} 
when $\rho-\chi_0 = 45^{\circ}$, and may reveal underlying structures that were confused by its presence.  Thus, the images at $\rho = \chi_0$ and $\rho = \chi_0 + 45^{\circ}$ will be most informative---but since we do not know what $\chi_0$ is for any given structure, we have to examine the full range of $\rho$ values.  This is really nothing new; since $Q$ and $U$ are an orthogonal basis, one may contain structures that are not present in the other.  However, what is new here is that by exploring a range of $\rho$ values, structures that were previously present in both $Q$ and $U$  can now be eliminated when the appropriate value of $\rho$ is used.

The most straightforward use of this technique is to examine a series of images at a discrete set of $\rho$ values. In practice, sampling $\rho$ every $\sim$10 degrees proves sufficient for initial exploration. Often, the disappearance of a specific structure at the relevant $\rho$ is what reveals other underlying polarized features. 

\subsection{An Illustrative Model: Multiple Polarized Components}

We present a series of S($\rho$) images here for a simple model of a multi-component  polarized structure. A large Gaussian component at fixed polarization angle 0$^{\circ}$ is overlaid with two series of fainter narrow Gaussians oriented at {\em position angles} +45$^{\circ}$  and $-$30$^{\circ}$, with {\em polarization angles} of 45$^{\circ}$ and $-$45$^{\circ}$, respectively.   In this particular case, the features are, therefore, isolated into the $Q$ Stokes image for the large Gaussian, and the $U$ Stokes image for the narrower Gaussians. 

\begin{figure}[H]
\includegraphics[width=8 cm]{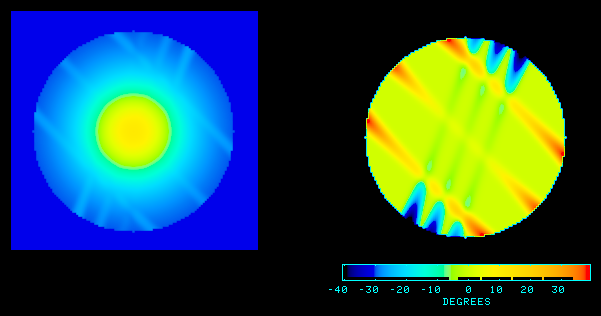}
\caption{{Model of overlapping Gaussian components as described in the text. \textbf{Left}: Polarized intensity;  \textbf{Right}: Polarization angle.}}
\label{stripes}
\end{figure}

\begin{figure}[H]

\includegraphics[width=8 cm]{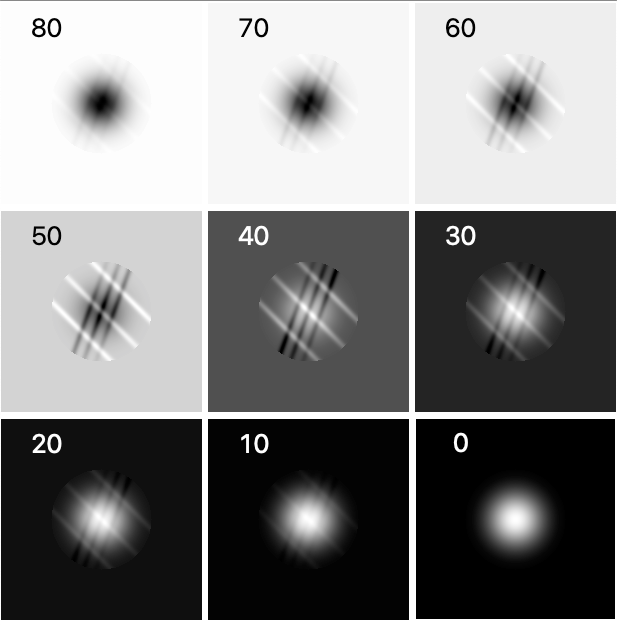}

\caption{Frames are each  $S(\rho)$ images of the simple model of multi-component polarized emission, as described in the text. The corresponding value of $\rho$ listed in the upper left. The greyscale is automatically set to cover the range in each frame, from black=minimum to white=maximum.  $S(\rho)$ can be either positive or negative; the background in each frame is at zero intensity.} 
\label{stripegrid}
\end{figure}

Figure \ref{stripes} shows the polarization intensity and polarization {angle}. From the polarized intensity map, the narrow gaussians are visible around the periphery, but it is not clear whether they {extend} all the way to the center.  However, the polarization angle image suggests that they do, but are perhaps considerably weaker near the center.  Around the periphery of the image, the polarization angle varies from +45$^{\circ}$ to $-$45$^{\circ}$, while at one quarter of the radius, the polarization angle varies from only +3$^{\circ}$ to $-$3$^{\circ}$.  Although this is  true, and what an observer might report, it is a misleading characterization of the underlying source structure, which we can see in the S($\rho$) images in Figure \ref{stripegrid}.  Note that in the frames where the large component has been greatly reduced (at $\rho$ values of  40 and 50), the narrow Gaussians are seen to be strong as they cross the center, not weak as might be inferred from the small variations in polarization angle observed near the center.  If this had been a real observation, then the $Q$ and $U$ images would look like any pair of images separated by 45$^{\circ}$, so whether or not the different features would have been well isolated is a matter of chance. In practice, in examining the $S(\rho)$ gallery, the astronomer has to make a scientific judgement as to whether newly revealed structures are significant, random fluctuations, instrumental effects, etc., just as they would with any image.

\subsection{An Illustrative Model---Polarization Angle Gradients}
\begin{figure}[H]

\includegraphics[width=8 cm]{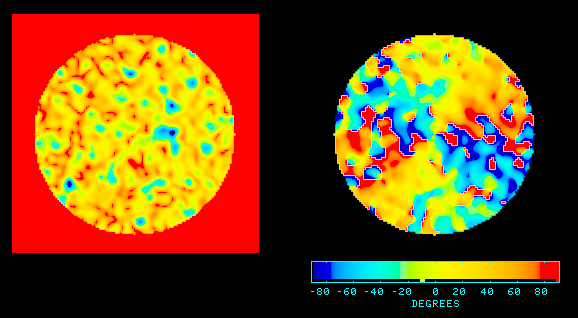}

\caption{Simple model of azimuthally varying polarization angle plus noise, as described in text. (\textbf{Left}) Polarized intensity (\textbf{Right}) Polarization angle.}
\label{azimuth}
\end{figure}

Our second example illustrates the power of polarization tomography to see subtle gradients in polarization angle as a function of position, by converting the changes into a time sequence, which is often easier to recognize.  We take all of the S($\rho$) images and visualize them as a movie.   In this model, we add a polarized signal of constant amplitude and variable polarization position angle $\chi(a)$ = $a+45\deg$, where $a$ is the azimuthal angle.  Random noise is then added with an rms amplitude equal to the polarized intensity. The resulting polarized intensity and polarization angle images are shown in Figure  \ref{azimuth}.   At a signal-to-noise ratio of 1, the polarization angle map appears to have an azimuthal dependence, although an observer might express caution with that conclusion.   The movie of S($\rho$) frames, Figure S1,  
makes the azimuthal dependence very clear. (S1 and S2, below,  are animated gifs, available under Ancillary Files on the arXiv download page for this preprint.)

 \section{Cygnus A}
Highly detailed polarization maps of the well-studied radio galaxy Cygnus A were presented by \citet{cyga}.  We carried out a polarization tomography analysis of the 0.2'' 
resolution image of the western lobe at 8~GHz to illustrate the application of the technique to real data. No correction has been made for Faraday rotation, so the behavior of the polarization angles reflects both Faraday effects and intrinsic magnetic field structure.

\begin{figure}[H]

\includegraphics[width=13cm]{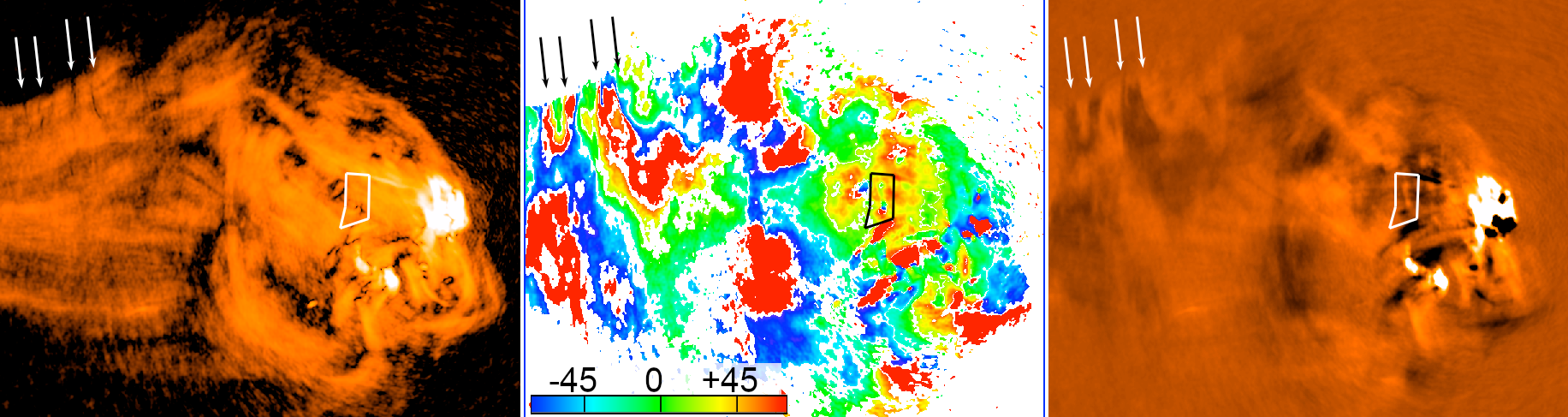}

\caption{Western lobe of Cygnus A at 8 GHz. (\textbf{Left}) Polarized intensity.  (\textbf{Middle}) Polarization angle. (\textbf{Right}) Tomography frame S($\rho=75^{\circ}$).   Arrows and the polygon refer to interesting features to look at in the tomography movie.} 
\label{cyga}
\end{figure}

Figure \ref{cyga} shows the polarized intensity and angle of the western lobe {in the left and middle panels, respectively, while the right panel shows a single frame from the S($\rho$) gallery.}  The polarized intensity shows a chaotic mixture of filaments behind the hot spot;  further downstream the filaments tend to align along the lobe.  By contrast, vertical bands are seen in the polarization angle, and the places where the gradient in angle is very strong lead to darker, depolarization filaments, such as those indicated by arrows.   The polygon encloses an area of no particular significance in either polarized intensity nor polarization angle.

 The tomography movie is shown in Figure S2. The movie first shows the frames in terms of increasing $\rho$, 0--180$^{\circ}$, and then reverses, as in S1.  The amount of detail appears at first overwhelming, and indeed there is much to learn from the movie which is beyond the scope of this paper.  However, we call your attention first to the locations indicated by the arrows, where Faraday rotation has caused vertical striping in the polarization angles.  The geometry of those gradients is visible in the movies; features such as pairs of dark (or light) bands which appear to expand and contract represent local minima and maxima in polarization angle, with the same gradient in both directions.  Other patterns of gradients, illuminating the Faraday and intrinsic field geometry, can be seen along filaments, e.g., in the tangled region behind the hot spot.   

The polygon illustrates a simple case where a distinct polarized feature becomes visible in the tomography images which is not clear from the polarized intensity map.  This small polarized filament does not have a single polarization angle, but the continuity of the structure in the movie allows it to be identified.

\section{Conclusions}

Polarization tomography is a simple, complementary tool to explore the rich detail available in polarization maps. It allows identification of features not visible in polarized intensity and polarization angle maps, especially due to interference with other polarized features. It also enables subtle patterns to be recognized in the polarization angle structure which would otherwise escape our attention.  It should be in widespread use.



\vspace{6pt} 


\supplementary{Figures S1 and S2 are animated gifs, and  are available under Ancillary Files at the arXiv download page.} 

\authorcontributions{Methodology, D.K., L.R.; Data analysis, L.S.; Writing---original draft preparation, L.R.; Writing---review L.S. All authors have read and agreed to the published version of \mbox{the manuscript.}}

\funding{This research was funded in part by U.S. National Science Foundation grant AST93-18959 to the University of Minnesota. We should have published it decades ago.}

\institutionalreview{Not applicable}

\informedconsent{ Not applicable}
%

\dataavailability{Not applicable}

\conflictsofinterest{The authors declare no conflicts of interest.}





\end{paracol}

\reftitle{References}

\end{document}